# Similarity Is Not Enough: Tipping Points of Ebola Zaire Mortalities

J. C. Phillips

Dept. of Physics and Astronomy, Rutgers University, Piscataway, N. J., 08854

## Abstract

In early 2014 an outbreak of a slightly mutated Zaire Ebola subtype appeared in West Africa which is less virulent than 1976 and 1994 strains. The numbers of cases per year appear to be ~ 1000 times larger than the earlier strains, suggesting a greatly enhanced transmissibility. Although the fraction of the 2014 spike glycoprotein mutations is very small (~3%), the mortality is significantly reduced, while the transmission appears to have increased strongly. Bioinformatic scaling had previously shown similar inversely correlated trends in virulence and transmission in N1 (H1N1) and N2 (H3N2) influenza spike glycoprotein mutations. These trends appear to be related to various external factors (migration, availability of pure water, and vaccination programs). The molecular mechanisms for Ebola's mutational response involve mainly changes in the disordered mucin-like domain (MLD) of its spike glycoprotein amino acids. The MLD has been observed to form the tip of an oligomeric amphiphilic wedge that selectively pries apart cell-cell interfaces via an oxidative mechanism.

**Introduction**  Protein amino acid sequences (aas) are rich in information, especially when combined with structural data. There are many Web-based tools for analyzing aas, but by far the most utilized is BLAST (**B**asic **L**ocal **A**lignment **S**earch **T**ool), which compares two given sequences, or searches for sequences similar to a given sequence. The original BLAST paper [1] was the most highly cited paper published in the 1990s. Ebola filoviruses exhibit ~ 30% spike glycoprotein (GP) subtype aas differences, with wide variations in virulence (from nearly 90% to nearly 0% mortality) [2]. Although the Marburg species has only 30% GP similarity to Ebola, its length and viral morphology are similar, and it also has high mortality levels ~ 50% [3,4]. The static GP domain structure of the most virulent Ebola subtype, Zaire or ZEBOV, whose structure bound to an antibody is shown in Fig. 1 of [5], is the basis of many studies [6-11]. The molecular basis for explaining the widely differential pathogenicity of the Ebola and Marburg filoviruses, which depends on multiple mechanisms involving many steps from molecular membrane penetration to selective disruption of cell-cell binding, remains elusive [2,7].

In 2014 a new GP ZEBOV* strain appeared, with apparently much increased transmissibility, and a recently estimated WHO mortality reduced from 90% to ~ 30%. Heroic efforts have sequenced GP ZEBOV*, which differs (BLAST) from the 1976 GP ZEBOV strain by only 3% [12]. Sequential similarity differences up to 60% normally do not alter structural homologies (folds), so conventional structure-function methods are uninformative here. Indeed, the surviving authors of [12] note that their data alone "do not address whether these [aas] differences are related to the severity of the outbreak."

Bioinformatic scaling had previously shown inverse correlated trends in virulence and transmission in N1 (H1N1) and N2 (H3N2) influenza spike glycoprotein strains [13]. These trends appeared to be related to various external factors (migration, cold and crowded conditions, and vaccination programs). As these conditions worsen, a "tipping point" [technically a thermodynamic critical point] can be reached, beyond which a pandemic occurs, its most famous example being the 1918 H1N1 world-wide influenza pandemic (3% mortality, 50 million deaths). There are collective social tipping points [14], which have become more prevalent as the Internet has facilitated "viral" information transmission.

For self-organized networks very near thermodynamic equilibrium there is a general theory of critical tipping points which appears to be relevant to the ZEBOV* outbreak [15]. Studies of toy models had shown that thermodynamic functions near critical points have self-similar (power law) behavior describable by fractals [16,17]. Proteins are compacted into globules, and a landmark bioinformatic paper identified the twenty amino-acid specific fractals MZ that describe hydropathic compaction forces [18]. These twenty fractals are associated with the differential geometry of solvent-accessible Voronoi surfaces centered on each amino acid. While the differences between the MZ hydropathic parameters discovered bioinformatically and the standard KD water-air enthalpy differences [19] are small [correlation(MZ,KD) = 0.85], it is just such small differences that become important near a thermodynamic critical point [20]. A striking example is β amyloid aggregation, responsible for Alzheimer's disease, where the MZ scale is nearly twice as effective as the KD scale in identifying molecular β sandwich structure, given only the β amyloid aas as input [21].

Two GP mechanisms have been suggested as possible origins of large Zebola subtype virulence variations, different folds (Ebola Zaire is the only structure known) [5], and different flexibility [7]. Ebola subtypes share 70% sequence similarity, and subtypes with more than 40% similarity generally have similar folds, so the fold explanation is unpromising [22]. A characteristic feature of filoviruses is a large, disordered, flexible and highly glycosylated mucin-like domain (MLD, 305-485), which is a natural candidate for disruption of cell-cell binding [7,11]. There are large differences between Ebola subtypes in the MLD, and strong similarities outside the MLD. This is shown in Fig. 1, which presents the results of a standard BLAST similarity analysis of the full (GP1, GP2) aas differences of ZEBOV and the least virulent subtype REBOV, using running windows. From this plot one suspects that the virulence differences between the Ebola subtypes arise from differences in their disordered MLD, but the elastic mechanism for these differences is unclear.

**Results and Discussion** Given a bioinformatic scale $\psi(aa)$, from Fig. 1 it is clear that the subtype functional differences associated with the 305-485 MLD occur on a large scale, which we have taken to be defined by a sliding window of width W = 115. The resulting $\psi(aa,115)$ profiles of the three subtypes with variable virulences (ZEBOV, 80-90% mortality, SEBOV, 40-60% mortality, and REBOV, ~ 0% mortality) are shown in Fig. 2. With the MZ scale, hydroneutral is near 155, so the ZEBOV profile consists of two hydroneutral peaks separated by an elastic hinge associated with the deep V-like hydrophilic minimum of the disordered MLD. The disordered structure of the MLD is highly variable (Fig. 1) and may be responsible for most of the widely differential pathogenicities. The most virulent subtype ZEBOV shows a deep, nearly smooth "V" minimum in the MLD near 420, while the least virulent REBOV subtype shows a similar minimum partially stabilized by a wide secondary hydrophobic peak (SHP) centered near 343. The intermediate SEBOV subtype has a similar stabilizing secondary hydrophobic peak centered near 437. These SHP are not resolved using BLAST (Fig. 1), and thus aas similarity alone cannot explain the differences in subtype mortalities.

These SHP stabilize the soft MLD hydrophilic ZEBOV V hinge, which becomes a strong candidate for explaining quantitatively the very strong ZEBOV virulence. One can also suppose that the unencumbered ZEBOV deep, nearly smooth "V" minimum near 420 enables the ZEBOV GP to function most effectively as an amphiphilic wedge that pries apart cell-cell interfaces [10,11,23]. This mechanism, which is obvious from Fig. 2, explains the haemorrhagic differences between Ebola subtypes [2]. A more detailed discussion of profiling differences for the three subtypes is given elsewhere [24], which includes Ebola domains adjacent to the MLD.

The present discussion has been most successful because it employs bioinformatics aas methods, which go further than similarity aas analysis alone [12]. The bioinformatics methods are based on analyzing thousands of protein aas, and identify universal fractal properties of proteins based on aa-dependent surface differential geometry. They are more accurate than enthalpy-based hydropathic scaling [19], which is based on water-air unfolding, an approximating process that does not occur in vivo. However, the underlying importance of hydropathic forces is so great that the SHP in Fig. 2 also occur when the KD scale is used instead of the MZ scale, but the resolution of the SHP structure is weaker (not shown here). Similarly better MZ than KD results were obtained for β amyloid aggregation [21].

Predicting fractal tipping points is an unsolved problem that has attracted much interest, especially in financial sectors [25,26]. How should we analyze the very small 3% GP mutation differences between ZEBOV and ZEBOV*? The study of the very small mutational differences involved in ZEBOV* requires special scaling methods to identify the domains where the correspondingly small critical changes have occurred. We examined GP profile differences between ZEBOV and ZEBOV* using the MZ and KD hydropathic scales [13,17,18] and a new thermodynamic bioinformatic scale FTI that recognizes incipient β strands associated with large hydrogen bond densities [27]. The FTI scales (Table 2 of [27]) include differences between exposed surface and buried residues, and the βexposed scale GP profiles exhibit the largest Z-Z* differences, as shown in Fig. 3. The FTI βexposed scale could be the most appropriate way to examine incipient secondary order (which should be weak and near the protein surface) in proteins with functionally important disordered domains, like the MLD. We discuss these differences using the domain structural diagram of Fig. 1a from [3].

The most obvious difference is the hydrophilic tilt [21,23] of the V profile of the MLD away from its C side 460-560 towards its N side 240-450. This could explain the reduced virulence. The second difference is the stiffening of the signal protein between 100 and 150. This could explain the enhanced transmissibility, as signaling proteins could readily combine to form the observed oligomeric MLD fist [28, 10,11]. These are small shifts, but near a tipping point they can have large effects. The small shifts are invisible using similarity methods, which cannot separate virulence and transmissibility.

This leaves an interesting question: is there a thermodynamic mechanism that would explain the inverse correlation between virulence and transmissibility that is clear for Z-Z* differences, and also for neuraminidase (GP influenza) evolution since vaccination programs began in 1945 [13]? We may suppose that there is a delicate thermodynamic balance between virus and antibodies. Initially external factors (in the Sierra Leone case, increasing lack of pure water in rural areas) tilt the balance in favor of the virus, but as the human network responds, the virus encounters increasing antibody resistance.   The viral network responds by mutating to increase its transmissibility, but it can do this best by decreasing virulence.   The 1918 extremely virulent influenza strain had disappeared by 1920, for reasons unknown.   Here with ZEBOV* it appears that we may have identified GP mutations in the MLD spike protein tip that decrease virulence.

At this time little is known about the factors responsible for ZEBOV* mortality, but many physicians' observations, such as the "light bulb" phenomenon, are suggestive of a "tipping point" [29].   Here we have analyzed the most relevant information at the molecular level, the GP aas [12].   Our influenza analysis had predictive character [13], and the present analysis may be tested against mutations of future ZEBOV* strains.   Readers who are encountering viruses for the first time will find an excellent discussion of viral mechanisms and spike glycoproteins in [30].

**Methods**   The calculations described here are very simple, given the scaling tables of [20] and [26].   They are most easily done on a spread sheet, such as a dedicated EXCEL macro.   The one used in this paper was built by Niels Voohoeve and refined by Douglass C. Allan.

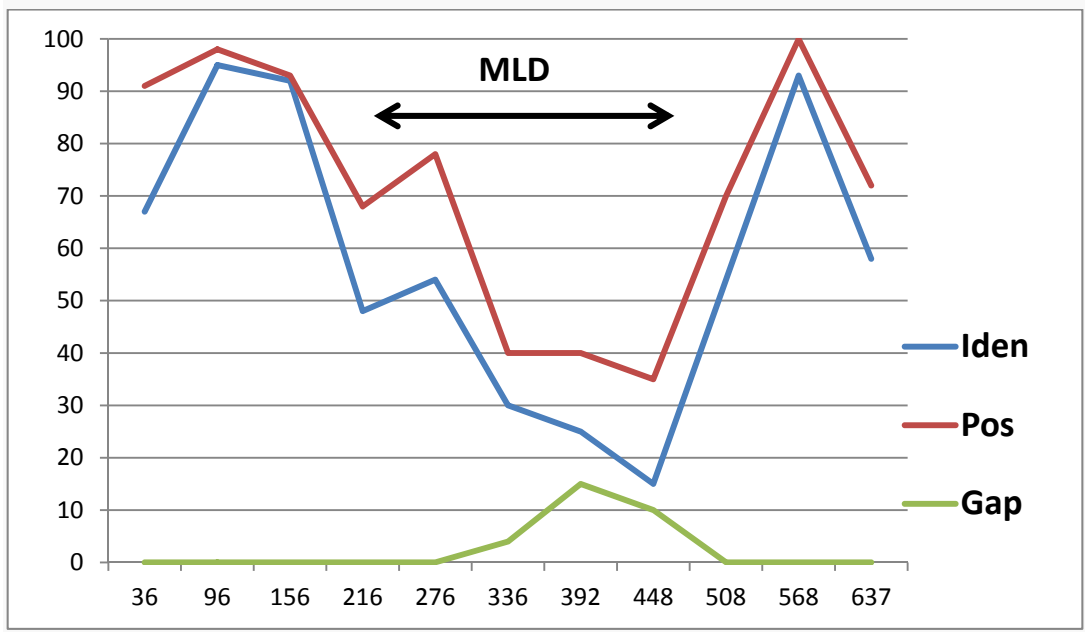

Fig 1. Running BLAST differences (Identity, Positives and Gaps) between GP ZEBOV and REBOV (Uniprot O11457 and Q66799). Similarity differences below 40% suggest structural differences in the MLD. The values shown refer to (roughly) W = 60 windows centered on the indicated sites, for instance, 216 is calculated for 187-246.

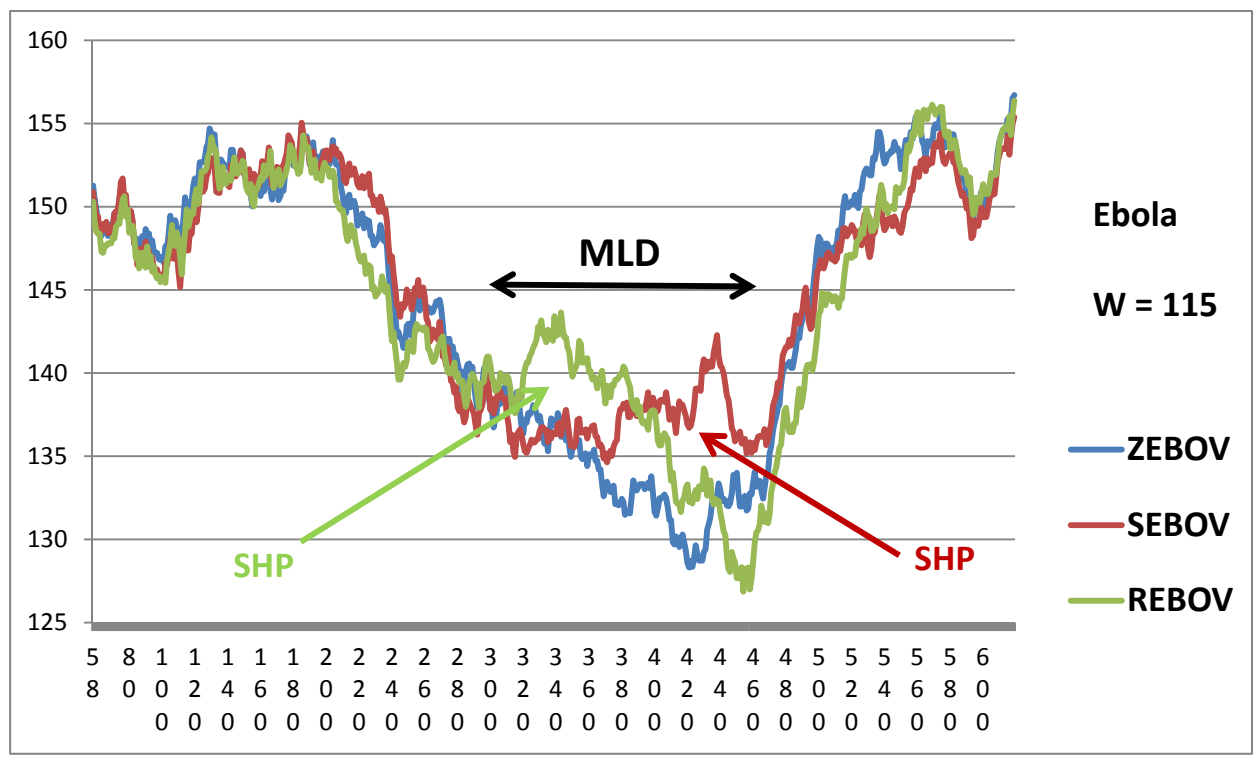

Fig. 2. Profiles of ψ(aa,115) for three Ebola GP subtypes. The MLD lies in the central hydrophilic valley, between the receptor domain plateau centered on 160, and the transmembrane anchor near 570. Secondary hydrophobic peaks for SEBOV and REBOV subtypes are marked.

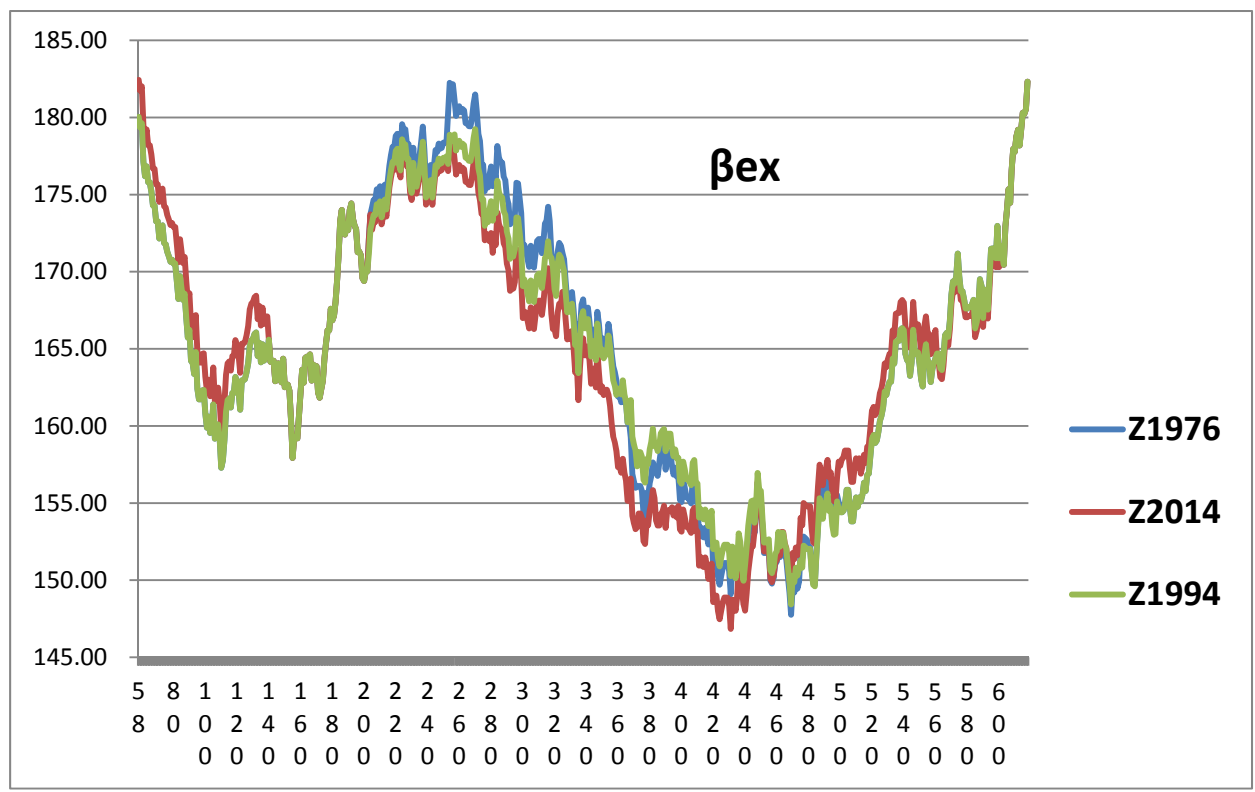

Fig. 3. Profiles of GP ZEBOV 1976 and 1994, sequences from Uniprot, and ZEBOV* from PRJNA257197. There are two large differences, from 250 to 550, and from 110 to 140. Here the FTI β exposed scale from [26] has been multiplied by 200 and transformed using a sliding window of width W = 115.